\documentclass[reprint,superscriptaddress, amsmath,amssymb, aps, pra, ]{revtex4-2}

\usepackage[usenames, dvipsnames]{color}

\usepackage{hyperref}
\hypersetup{
    colorlinks=true,
    linkcolor=Blue,
    citecolor=Blue,
    filecolor=Blue,      
    urlcolor=Blue,
    pdftitle={Spinor droplets},
    }
\usepackage{graphicx}
\usepackage{physics}

\newcommand{\PsiCreate}[1]{\hat{\Psi}_{#1}^{\dagger}(\mathbf{x})}
\newcommand{\PsiAnn}[1]{\hat{\Psi}_{#1}(\mathbf{x})}
\newcommand{\akm}[2]{\hat{a}_{\mathbf{#1},#2}}
\newcommand{\akmDagger}[2]{\hat{a}^{\dagger}_{\mathbf{#1},#2}}
\newcommand{\FExpected}{\langle \mathbf{F}\rangle}
\newcommand{\Fmm}[1]{\mathbf{F}_{#1}}
\newcommand{\DkDagger}[1]{\hat{D}_{\mathbf{#1}}^\dagger}
\newcommand{\Dk}[1]{\hat{D}_{\mathbf{#1}}}
\newcommand{\FkDagger}[1]{\mathbf{\hat{F}}_{\mathbf{#1}}^\dagger}
\newcommand{\Fk}[1]{\mathbf{\hat{F}}_{\mathbf{#1}}}
\newcommand{\ek}{\epsilon_{\mathbf{k}}}
\newcommand{\cbar}[1]{c_{#1}}

\def\xBold{\mathbf{x}}
\def\U0{\tilde{U}(0)}

\def\a-kd{a^{\dagger}_{-k}}

\def\b-kd{b^{\dagger}_{-k}}

\def\sumk0{\sum_{\vec{k} \neq 0}}

\begin{document}

\title{Spinor Boson Droplets Stabilized By Spin Fluctuations}

\author{T.A. Yo\u{g}urt}
 \email{ayogurt@metu.edu.tr}
 \affiliation{%
Department of Physics, Middle East Technical University, Ankara, 06800, Turkey\\
}%
\author{A. Kele\c{s}}
 \affiliation{%
Department of Physics, Middle East Technical University, Ankara, 06800, Turkey\\
}%

\author{M.\"O. Oktel}
\affiliation{Department of Physics, Bilkent University, Ankara, 06800, Turkey}

\date{\today}

\begin{abstract}
Self-trapped droplets stabilized by quantum fluctuations have been experimentally realized in dipolar gases and binary Boson mixtures. We propose spinor Bose gases as another candidate for droplet formation in this work. For spin-1 gas, we find that spin fluctuations give a dilute but self-trapped state for two different order parameters where the mean-field picture predicts collapse. A polar droplet phase can be stabilized by spin fluctuations for both antiferromagnetic and ferromagnetic spin-dependent coupling. An antiferromagnetic droplet phase can be stabilized similarly with a negative quadratic Zeeman shift. Furthermore, the beyond mean-field energy of the system depends on the quadratic Zeeman coupling, which provides a mechanism to tune the droplet formation and its density. We discuss the parameters necessary for the experimental realization of such spinor droplets.  
\end{abstract}

\maketitle

\section{\label{sec:level1} INTRODUCTION}

Bose-Einstein condensates (BECs) have generally been modeled by the Gross-Pitaevskii mean-field (MF) approach to explain various experimental observations. Such MF theories of interacting BECs provide both a reasonable quantitative agreement and a qualitative understanding of complex phenomena, including collapse or expansion dynamics of the condensate, collective modes, bright and dark solitons, and shift of critical temperature due to interactions \cite{pitaevskii2016bose,pethick2008bose,1996_RPL_Ketterle,1999_RMP_Stringari,2004_PRA_Aspect,2020_PRL_Stringari,2011_PRL_Hadzibabic}. The Bogoluibov theory of weakly interacting Bose gas includes additional beyond MF processes such as the scattering of two atoms in the condensate to states with momenta $\mathbf{p}$ and $-\mathbf{p}$. The calculated corrections to the energy dispersion and the non-condensate depletion due to these quantum fluctuations are experimentally verified through the measurement of condensate excitations \cite{2008_PRL_Cornell,2011_PRL_Salomon}, which expose deviations from the MF theory. However, a more striking manifestation of the quantum fluctuations is the recently obtained by self-bound Bose droplets \cite{2015_Petrov_PRL,2018_Science_Tarruel_Mixture_Droplet,2016_PRL_Pfau_Dipolar_Droplet}.

The traditional repulsively interacting BECs are mechanically stabilized and prevented from expansion with confining potentials based on magnetic or optical traps \cite{pitaevskii2016bose}. For attractive interactions, higher densities are energetically favorable, and the collapse of the gas to a high-density non-trapped phase can only be prevented in a metastable state, such as a bright soliton \cite{2006_PRL_Bright_Soliton_Collapse}.    
The central point of novelty in the self-bound droplets is the use of the quantum fluctuations to establish mechanically stable BECs in a regime where the MF theories predict collapse. This stability can be achieved without violating the diluteness assumption for the gas, only if separate physical parameters control the mean-field interaction energy and the quantum fluctuation contribution  \cite{2015_Petrov_PRL}.

Recent experiments have realized two classes of ultracold droplets:  binary mixture droplets \cite{2018_Science_Tarruel_Mixture_Droplet,2018_RPL_Modugno_Mixture_Droplet,2021_PRA_Cornish_Mixture_Droplet}  and dipolar droplets \cite{2016_PRL_Pfau_Dipolar_Droplet,2016_Nature_Pfau}, both of which exhibit a tunable competition among distinct interactions and relatively weak Lee-Huang-Yang (LHY) corrections that can stabilize the residual MF energy. In dipolar droplets, the long-range and short-range interactions compete at the MF level, whereas in the binary mixture droplets, the interspecies and intraspecies short-range interactions combine to form two independent parameters, which individually control the MF and quantum fluctuation corrections. In the latter, if  the s-wave scattering lengths for intraspecies interactions are $a_{11}>0$, $a_{22}>0$ and the interspecies scattering length  is $a_{12}<0$,  the residual MF interaction is proportional to $n^2 \delta a$, where $\delta a = -|a_{12}| + \sqrt{a_{11}a_{22}}$. For negative $\delta a$, the MF energy favors higher densities and drive the system to collapse. The LHY interaction energy is of the form $(a_+ n)^{5/2} $, where $a_+>0$ is the effective scattering length for quantum fluctuations \cite{2015_Petrov_PRL}. The LHY term increases faster with density and prevents collapse. If the equilibrium density  does not violate the diluteness assumption of the Bogoliubov theory, an ultra-dilute, yet liquid-like self-bound droplet emerges. 

The experimental results of the binary mixture droplets in homonuclear \cite{2018_Science_Tarruel_Mixture_Droplet,2018_RPL_Modugno_Mixture_Droplet} and heteronuclear \cite{2021_PRA_Cornish_Mixture_Droplet} systems are in fair agreement with theories based on Gross-Pitaevskii approximation with local LHY corrections. However, a deeper understanding of such fluctuation stabilized states is desirable for two reasons. Firstly, better quantitative agreement with experiments is required  \cite{2018_Science_Tarruel_Mixture_Droplet,2020_PRL_Liu_Bosonic_Pairing}. Secondly, these systems can be used to test the validity of various theoretical proposals in quantum many-body physics. As an example, consider the unstable soft Bogoliubov modes in the theory of the binary mixture droplets \cite{2021_PRA_Gajda_Stability, 2020_PRL_Liu_Bosonic_Pairing}. While some recent work claim that these modes can be stabilized due to exotic many-body effects like bosonic pairing \cite{2020_PRL_Liu_Bosonic_Pairing} or beyond LHY contributions \cite{2020_PRB_Yin_Stability,2020_SPP_Ota_Stability}, other theories neglect them, claiming that the instability would be too slow to be observed in the experiment. Broadening the family of droplets \cite{2021_PRL_Petrov_Bubble, 2020_PRA_Mazzanti_SpinOrbitDroplet, 2021_Arxiv_Cui_Borromean_Droplet, 2021_PRL_Blakie_BinartMagneticDroplet, 2021_PRL_Santos_Dipolar_Mixture} by adding a new stabilization mechanism may lead to novel phenomena and further enhance droplet theories. 

In this paper, we consider spin-1 BEC gas as a candidate for self-bound droplet formation. We show that spin fluctuations can stabilize the polar and the antiferromagnetic phases of the spin-1 gas in the parameter regime for which the MF theory predicts collapse. Spinor gas s-wave scattering lengths cannot be changed using the standard Feshbach resonances, and the currently obtained spinor gases in Na and Rb are stable against density collapse within MF theory \cite{1998_Nature_Ketterle_Spinor,2004_PRL_Chapman_spinor,2004_PRL_Schmaljohann_Spinor}. However, there are proposals such as optical Feshbach resonances \cite{2015_PRA_Optical_Feschbach_Resonance,2018_Nature_Ott_Optical_Feshbach_Resonance}, which may provide new ways to tune the gas into the droplet regime in future experiments. 
As more and more atom species are cooled to ultracold temperatures, 
it is essential to investigate the hyperfine manifolds with unstable mean-field ground states and analyse the possibility of stable self-bound droplets. Whether obtained by controlling scattering lengths or through naturally occurring scattering lengths, 
a droplet stabilized with spin fluctuations offers additional tools to investigate the nature of the beyond-mean-field equilibrium, such as the quadratic Zeeman shift.

This paper is organized as follows. In Section \ref{sec:Bogoliubov Theory}, we summarize the Bogoliubov Theory of spin-1 gas, discuss possible mean-field magnetic orders and the stability when LHY corrections are introduced. In Section \ref{sec: Polar Spin-1 Droplet}, we develop the formulation of the spinor droplet within the polar spin-1 phase and present our numerical results. In Section \ref{sec:AF Spin-1 Droplet}, we derive the parallel formulation for the antiferromagnetic droplet phase. In Section \ref{sec:Experimental Discussion and Conclusion}, we discuss the experimental feasibility of spinor droplet formation and present our conclusions.

\section{\label{sec:Bogoliubov Theory} Spin-1 Gases: Bogoliubov Theory}

We consider a BEC of spin-$1$ atoms with s-wave interactions under an applied uniform and static magnetic field along the $z$-axis. We assume a low dipolar relaxation rate, which conserves the overall magnetization along the $z$-axis. We set this conserved magnetization to zero and drop the linear Zeeman terms from the Hamiltonian. Quadratic Zeeman energy $q$ is taken into account as in \cite{2013_RMP_Spinor_Kurn_Ueda,2010_PRA_Ueda_Spinor_LHY}. The Hamiltonian is given by 
\begin{eqnarray}
\hat{H} &&= \int d\xBold{} \  \PsiCreate{m} \left( -\frac{\hbar^2 \nabla^2}{2M} + qm^2 \right)\PsiAnn{m}\nonumber \\
&& + \frac{\cbar{0}}{2} \int d\xBold{} \ \PsiCreate{m} \PsiCreate{m'} \PsiAnn{m'} \PsiAnn{m} \\&&+ \frac{\cbar{1}}{2}\int d\xBold{} \  \PsiCreate{m} \PsiCreate{m'}\ \Fmm{m n} \cdot \ \Fmm{m' n '} \PsiAnn{n '} \PsiAnn{n}   \nonumber 
\label{spin1Hamiltonian}
\end{eqnarray}
where  $\PsiCreate{m}$ and $\PsiAnn{m}$ are creation and annihilation operators of spin-1 atoms with the magnetic quantum number $m$,
$\Fmm{mm'} = (F_{mm'}^x,F_{mm'}^y,F_{mm'}^z)$ are the set of spin-1 matrices, represented in the basis of $z$-axis eigenstates, and summation over $-1$, $0$ and $1$ is implied with repeated indices. Density and spin coupling constants, $c_0$ and $c_1$, are written as 
$\cbar{0} = (g_0 +2g_2)/3$ and $\cbar{1}=(g_2 -g_0)/3$. Here, $ g_0$ and $g_2$ are the 
bare coupling constants  of the $s$-wave collisions of two spin-1 bosons with total spin-$0$ and $2$, which are given in terms of  the corresponding scattering lengths $a_{0,2}$ as
$g_{0,2} = 4\pi a_{0,2} \hbar^2/M$. 
The bare coupling constants $g_{0,2}$ are renormalized using the standard $T$-matrix perturbation up to the second-order to remove the ultraviolet divergence in the beyond mean-field energy.
Note that both static magnetic fields and RF pulses can be used to adjust the strength of the quadratic Zeeman energy $q$ \cite{2004_PRA_Hirano_MagneticControl_q,2007_PRL_Let_RF_Quadratic}. 

Using  Bogoliubov theory, 
we replace the operators $\akm{0}{m}$ with the c-numbers $\sqrt{N_0} \tau_m$ and keep the terms with $\akm{k \neq 0}{m}$ and  $\akmDagger{k \neq 0}{m}$ up to the second order, where $N_0$ is the number of particles in the $\mathbf{k} = 0$ state, $\tau$ is the ground state order parameter in the spin-1 manifold
and $\hat a_{\mathbf{k},m}$ are the Fourier components 
$\PsiAnn{m} = V^{-1/2}\sum_{\mathbf{k}} \akm{k}{m} e^{i\mathbf{k}\mathbf{x}}$. 
Then, the effective Bogoliubov Hamiltonian for spin-1 Bose gas becomes \cite{2010_PRA_Ueda_Spinor_LHY}: 
\begin{eqnarray}
\hat{H}_{eff} = &&\frac{Vn^2}{2}(c_0 + c_1 \FExpected^2) + qN\langle F_z^2 \rangle \nonumber \\
&& + \sum_{\mathbf{k \neq 0}}  \Big \{ \left[ \ek - nc_1 \FExpected^2 + qm^2 - q\langle F_z^2 \rangle \right] \akmDagger{k}{m} \akm{k}{m} \nonumber\\
&&+ nc_1\FExpected \cdot \Fmm{mm'}  \akmDagger{k}{m} \akm{k}{m'} \nonumber\\ &&+ \frac{nc_0}{2}(2\DkDagger{k} \Dk{k} + \Dk{k}\Dk{-k} + \DkDagger{k} \DkDagger{-k}) \nonumber \\
&&+ \frac{nc_1}{2}(2\FkDagger{k} \Fk{k} + \Fk{k}\Fk{-k} + \FkDagger{k} \FkDagger{-k}) \Big \}  
\label{denklem}
\end{eqnarray}
where $\ek = \hbar^2 \mathbf{k}^2/2M$ is the free particle dispersion, $\FExpected \equiv \sum_{m,m'} \Fmm{mm'}\tau_m^*\tau_{m'}$ is the expectation value of the spin-1 order parameter, $\Dk{k} \equiv \sum_{m} \tau_m^* \akm{k}{m}$ and $\Fk{k} \equiv \sum_{m,m'} \Fmm{mm'} \tau_m^* \akm{k}{m'}$ are the density and spin fluctuation operators, respectively. 
The first line of \eqref{denklem} is the MF energy functional of the spinor BEC
\begin{eqnarray}
\frac{E_\mathrm{MF}}{V} = &&\frac{n^2}{2}(c_0 + c_1 \FExpected^2) + qn\langle F_z^2 \rangle 
\label{eq:mean_field_energy}
\end{eqnarray}
whereas all the other terms within summation constitute the quantum fluctuations. 
Unlike the Bose-Bose mixtures, the quantum fluctuations within the spinor gases involve not only pseudo-spin labeling of different components, but true spin fluctuations. 

\begin{figure}
    \centering
    \includegraphics{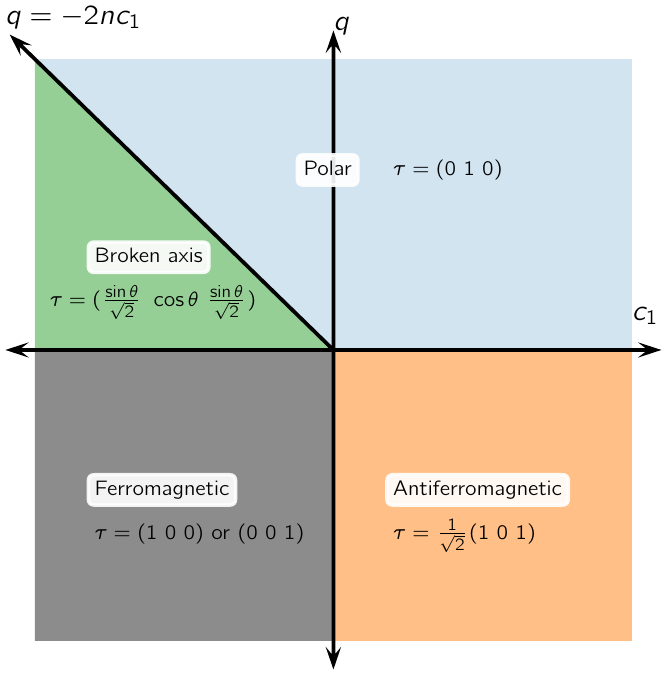}
    \caption{Schematic phase diagram of spin-1 Bose condensate as a function of quadratic Zeeman shift $q$ and spin-spin interaction $c_1$ for vanishing linear Zeeman energy $p=0$.}
    \label{fig:phase_diagram}
\end{figure}

The minimization of the spinor MF functional \eqref{eq:mean_field_energy} reveals the wide variety of magnetic orders and quantum phase transitions in spin-1 systems \cite{2013_RMP_Spinor_Kurn_Ueda}, which are briefly summarized here for completeness. See Fig.~\ref{fig:phase_diagram} for a schematic phase diagram. For $c_1 <0$ and $q<0$, the MF ground state is ferromagnetic with order parameters $\tau^\mathrm{F} = (1 \  0 \  0)$ or $(0 \  0 \  1)$ resulting in $\FExpected^2 = 1$ and $ \langle F_z^2 \rangle =1$. 
For $c_1 <0$ and $q>0$, the ground state depends on the strength of $q$: If $q>2|\cbar{1}|n$, the order parameter becomes $\tau^\mathrm{P} = (0 \  1 \  0)$ which is called the polar phase, and if $0<q<2|\cbar{1}|n$ 
the order parameter becomes $\tau^\mathrm{BA} = (\sin{\theta}/\sqrt{2} \  \cos{\theta} \  \sin{\theta}/\sqrt{2})$ with 
$\sin{\theta} = \sqrt{1/2 -q/(4|\cbar{1}|n)}$, which is named as the broken-axisymmetric phase. 
For $\cbar{1}>0$ and $q>0$, MF energy is again minimized with the polar order parameter $\tau^\mathrm{P}$.
Finally, for $\cbar{1}>0$ and $q<0$, the antiferromagnetic phase is obtained with order parameter $\tau^\mathrm{AF} = (1 \  0 \  1)/\sqrt{2}$. 

Let us now consider  mechanical instabilities for all the mean-field phases. In the ferromagnetic phase one can obtain 
the MF energy density as $n^2 (c_0 + c_1)/2 + qn$, which suggests a density collapse for $c_0 + c_1 \equiv g_2<0$. 
In the ferromagnetic phase, either $m=1$ or $m=-1$ spin state exist in a uniform condensate with total conserved magnetization along $z$-axis. This means two spin-1 bosons each having $m=1$ (or $m=-1$) can only scatter in a single collision channel with total spin $2$. Therefore, the ferromagnetic spinor gas acts like a single-component BEC with the well known LHY energy giving the following total beyond mean field energy
\begin{eqnarray}
\frac{E_0^F}{V} = qn + \frac{2\pi\hbar^2n^2}{M}a_2\left(1+\frac{128}{15\sqrt{\pi}} \sqrt{na_2^3}\right) 
\label{ferroGroundState}
\end{eqnarray}
which is written in terms of scattering length $a_2$ for convenience. It can be seen that the collapse in the MF level with $g_2\propto a_2<0$ cannot be prevented with the LHY term which comes with the same overall factor $a_2$. For this reason, one cannot obtain a stability mechanism due to the LHY energy for the spinor gas in the ferromagnetic phase.
We also note that LHY correction for the ferromagnetic order is independent of the quadratic Zeeman energy.
A similar analysis carried out for the broken-axisymmetric phase shows that droplet formation is not possible as mean field and LHY terms are controlled by the same parameter.

Consider the polar phase with $c_1 >0$ and $q>0$
(first quadrant in
Fig.~\ref{fig:phase_diagram}), 
which has the order parameter $\tau^P = (0 \ 1 \  0)$ and the MF ground state energy 
$E_{MF}^P = Vn^2c_0/2$ resulting in a density collapse for $c_0 <0$. Including the LHY corrections, the total ground state energy becomes for $c_1>0$ and $q>0$ \cite{2010_PRA_Ueda_Spinor_LHY}:
\begin{eqnarray}
\frac{E_0^P}{V} &&= \frac{n^2 c_0}{2} 
\left[ 1 + \alpha\sqrt{n c_0^3}\right] + \alpha n^2|c_1|\sqrt{n|c_1|^3} I(t),
\label{polarGroundState}
\end{eqnarray}
where $\alpha=16\sqrt{M^3}/15\pi^2 \hbar^3$, $t\equiv q/n|c_1|=|q|/n|c_1|$ since $q>0$, and
\begin{eqnarray}
I(t)  &&\equiv-\frac{15}{8\sqrt 2} \int_0^{\infty} dx \ x^2 \\
&&\times \left(x^2+t+1-\sqrt{(x^2+t)(x^2+t+2)} - \frac{1}{2x^2}\right).
\nonumber
\label{I1_numerical}
\end{eqnarray}
As shown in the Appendix, $I(t)$ can be approximated analytically as
\begin{eqnarray}
I(t)  \approx \frac{15\pi}{32\sqrt{2}} \left[ \sqrt{t+1} - \frac{1}{32} \frac{1}{(t+1)^{3/2}} \right].
\label{I1_analytical} 
\end{eqnarray}
to great accuracy which will be used in the following.

Crucially, beyond MF correction in the polar phase 
involves contributions from density and spin fluctuations, the terms with $\alpha$ that are proportional to $c_0$ and $c_1$ in \eqref{polarGroundState}, respectively, and it also depends on the quadratic Zeeman coupling $q$, in stark contrast with the ferromagnetic phase discussed above. 
In the limit $q=0$, $t\rightarrow 0$ and $I\rightarrow 1$, and $I$ monotonically increases with $q$. Thus spin fluctuations increase with quadratic Zeeman energy $q$, increasing the total ground state energy. 
In the ultra-low density limit, MF energy and density fluctuations scale with $n^2$, $n^{5/2}$, respectively, whereas spin fluctuations scale with $n^{2}$ due to $\sqrt{t+1}$ term in \eqref{I1_analytical}. Importantly, an instability initiated in the density channel with attractive interactions at the MF level, $ c_0<0$, can be countered with the quantum fluctuations in the spin channel, $ c_1$, which can be controlled with the Zeeman field $q$ whereas the fluctuations in the density channel are subleading.
This observation should be compared with quantum mechanical stabilization of Bose-Bose mixtures in Ref.~\cite{2015_Petrov_PRL} where a weak attractive interaction between the two components induces an MF-level instability, which is balanced by one of the two terms in the total LHY fluctuation. In contrast, the other LHY term gives a negligible ``soft-mode'' contributions that are routinely neglected in the literature of quantum droplets \cite{2015_Petrov_PRL,2018_RPL_Modugno_Mixture_Droplet,2019_PRA_Oktel,2019_NRP_Malomed,2020_RPP_Pfau,2021_PRA_Gajda_Stability}. In the following, we will also ignore these fluctuations and investigate the stability condition between the MF energy in the density channel and the LHY correction in the spin channel. 

\section{\label{sec: Polar Spin-1 Droplet} Polar Spin-1 Droplet}

\begin{figure*}[t]
    \includegraphics[width=0.4\textwidth]{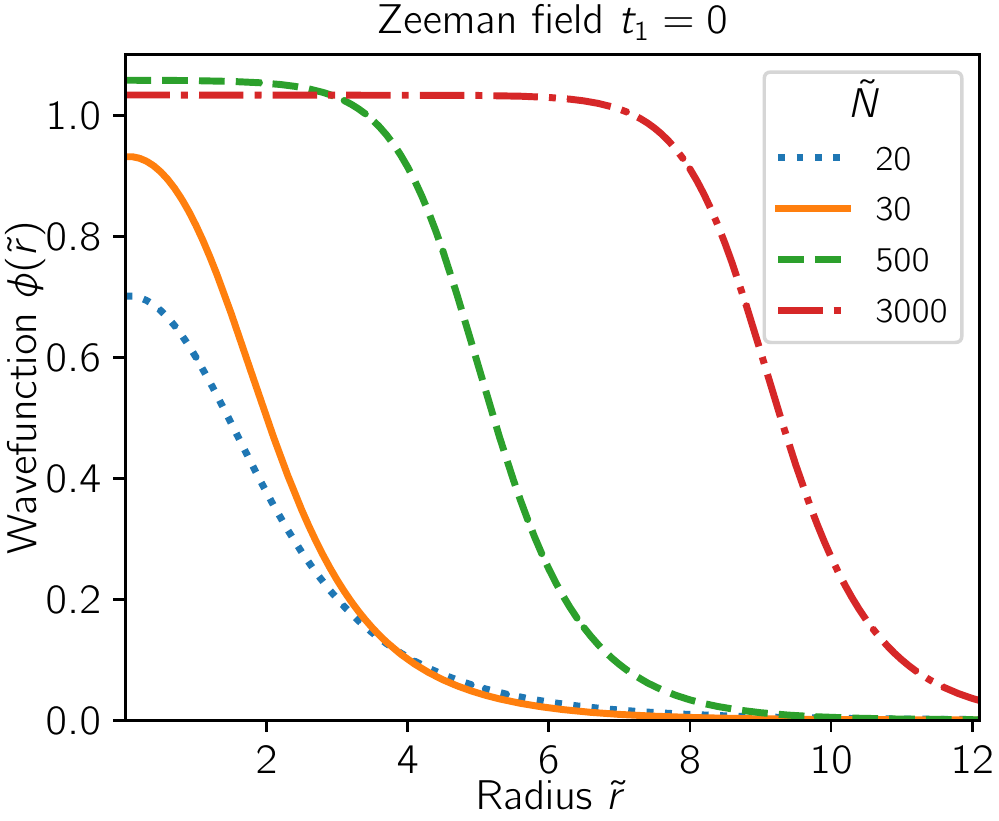}
    \includegraphics[width=0.4\textwidth]{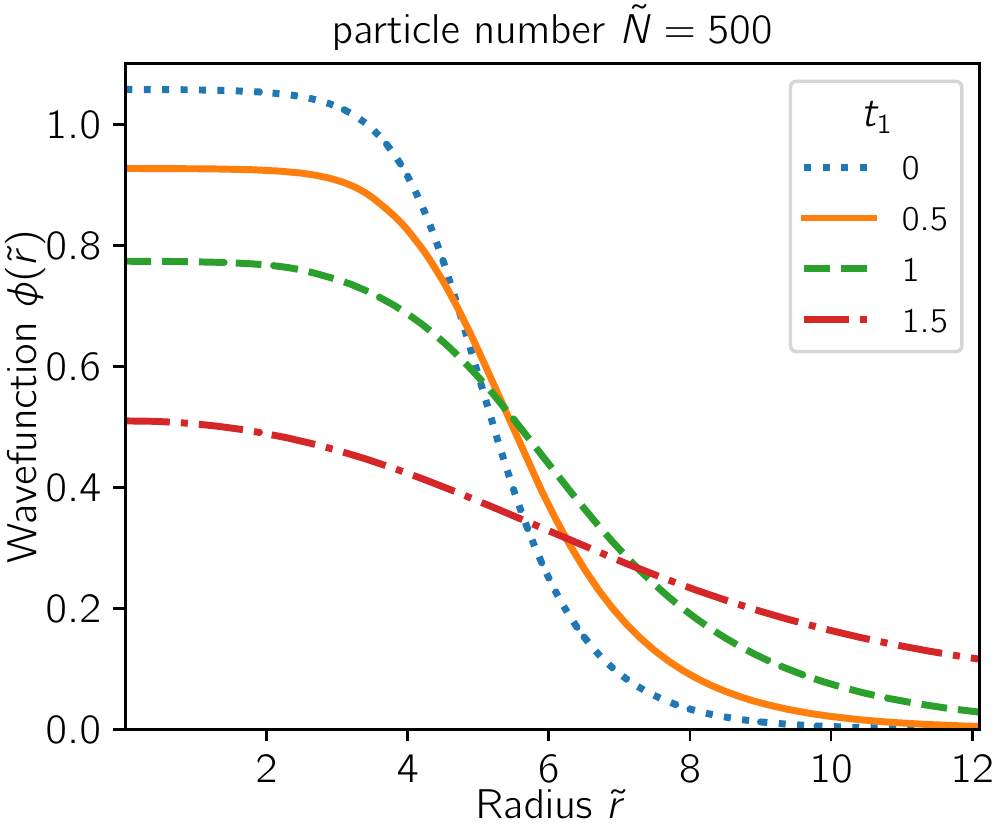}
    \caption{ \label{FigureDropletWavefunctions} The ground state wavefunctions of the spin-1 gas in polar phase for various values of particle number $\tilde{N}$ and Zeeman energy $t_1=q/n_1|\cbar{1}|$. (Left) The wavefunctions for $q=0$ (or $t_1=0$) and different values of  $\tilde{N}$. Below $\tilde N=20$ droplet is no longer self trapped. (Right) The wavefunctions for fixed $\tilde{N}=500$ and varying $t_1$, which shows self-bound droplet until a critical value of the quadratic Zeeman energy $q_c \approx 1.5$. The numerical calculation was done in the radial region $\tilde{r} \in [0,18]$. }  
\end{figure*}

For an infinite homogeneous spinor gas in the polar phase, the equilibrium between the MF and the spin fluctuation LHY can be calculated from the condition of vanishing pressure. Using the thermodynamic identity $P = -(\partial E/\partial V)_N$, we obtain 
\begin{eqnarray}
P &&= \frac{n^2c_0}{2} +
\frac{\alpha n^2|c_1|}{2} 
\sqrt{n|c_1|^3}f(t) 
\end{eqnarray}
where $f(t) \equiv 3I(t) - 2tI'(t)$ with  $f(0) = 3$. Setting this expression to zero gives the condition for the equilibrium density $n_0$ as 
\begin{eqnarray}
n_{0} = \frac{|\cbar{0}|^2}{\alpha^2 |\cbar{1}|^5 f^2(t_{0})},
\label{eq:n0_polar}
\end{eqnarray}
which is an implicit equation with 
$t_{0} = q/n_{0}|\cbar{1}|$. Here, $n_{0}$ also approximates the value of the saturation density in finite droplets for which the kinetic energy is negligible. Since $f(t)$ is a monotonically increasing function of $t$, the equilibrium density decreases with increasing quadratic Zeeman energy $q$.  Increasing $q$ provides a stronger LHY energy from spin fluctuations in the polar phase, and the equilibrium with the negative MF energy is reached at lower densities. 

Let us study the feasibility of a finite spinor gas in the polar droplet 
phase more quantitatively. For a polar spinor with wavefunction $\Psi(\mathbf{r})=
\psi(\mathbf{r})\tau^{P}$, we define the total energy functional as
\begin{eqnarray}
\mathcal{E}[\psi^*,\psi] &=& \frac{\hbar^2}{2M}|\nabla\psi|^2 
\nonumber\\
&+&\frac{c_0}{2}|\psi|^4
+\alpha |c_1|^{5/2}|\psi|^5
I\left[\frac{q}{|c_1||\psi|^2}\right]
\end{eqnarray}
and parametrize the wavefunction $\psi(\mathbf{r}) = \sqrt{n_1}\phi(\mathbf{r})$ with $n_1=|c_0|^2/9\alpha^2|c_1|^5$ obtained from \eqref{eq:n0_polar} in the limit $q=0$. We minimize the total energy in the grand canonical ensemble $E=\int d^3\mathbf{r}\mathcal{E}[\psi^*,\psi]-\mu N$ where the chemical potential is fixed by the total number of particles $N=\int d^3\mathbf{r}|\psi|^2$. The resulting modified GP equation is given by
\begin{eqnarray}
\tilde{\mu} \phi &&= -\frac{1}{2}\tilde{\nabla}^2\phi -3|\phi|^2\phi \nonumber \\
&&+\left[ \frac{5}{2} I \left( \frac{t_1}{|\phi|^2} \right) |\phi|^3 - I'  \left( \frac{t_1}{|\phi|^2} \right) t_1\  |\phi| \right]\phi,
\label{modifiedGPSpinor}
\end{eqnarray}
which is written in dimensionless form $\tilde{\mathbf{r}}=\mathbf{r}/\xi$, with $\xi = \sqrt{3\hbar^2/M |\cbar{0}|n_1}$ and $t_1 = q/n_1|c_1|$. Dimensionless chemical potential $\tilde\mu$ is determined from total particle number using $\tilde N = \int d^3 \tilde{ r}|\phi(\tilde{\mathbf{r}})|$ and related to the total number in the droplet as $\tilde N = N/n_1\xi^3 $. 

In the limit of vanishing Zeeman energy, $q=0$, Eq.(\ref{modifiedGPSpinor}) reproduces the modified GP equation of Ref.~\cite{2015_Petrov_PRL}. At this point, it is helpful to examine the correspondence between the polar spinor and binary mixture droplet more closely: The  MF instability condition $\cbar{0}<0$ in polar spinor corresponds to $\delta g<0$ in binary mixture, where $\delta g=g_{12}+\sqrt{g_{11}g_{22}}$ is written in terms of inter- and intra-component coupling constants  $g_{12}$ and $g_{11}$-$g_{22}$, respectively. The quantum fluctuations stabilizing this instability $n^{5/2}|\cbar{1}|^{5/2}$ in polar phase corresponds to the out-of-phase term
$n^{5/2}a_+^{5/2}$ in binary mixtures.
The ``soft modes'' that are neglected $n^{5/2}|\cbar{0}|^{5/2}$ in the polar phase corresponds to the in-phase term 
$n^{5/2}a_-^{5/2}$ of binary mixtures.

\begin{figure}
\includegraphics[width=0.4\textwidth]{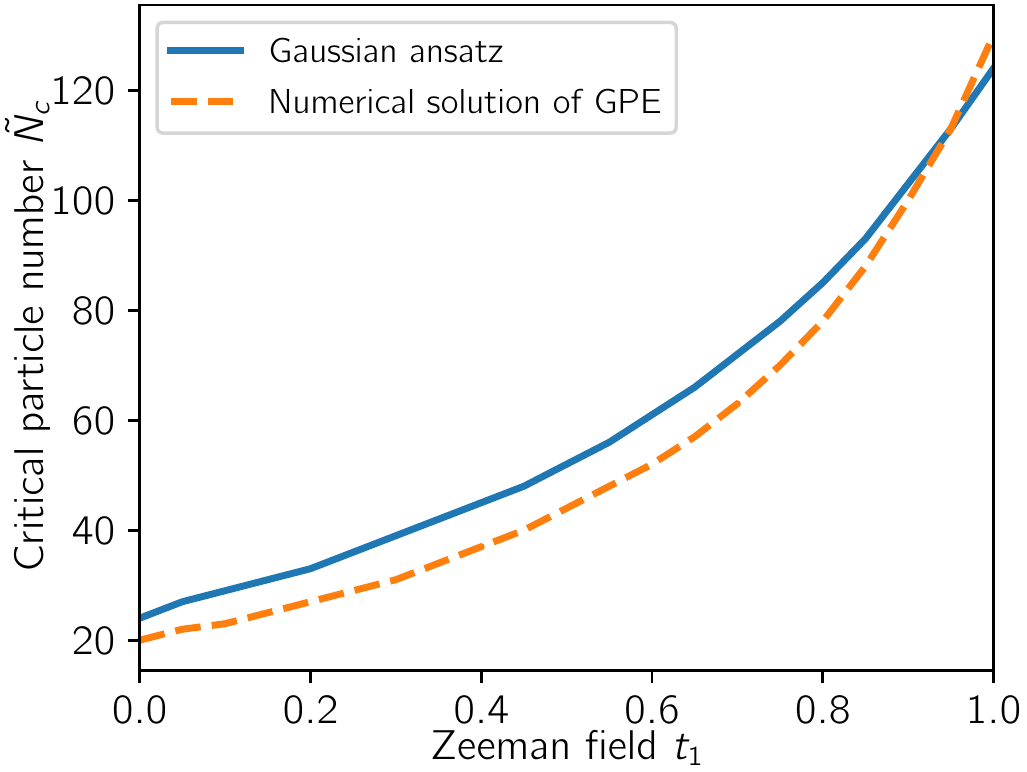}
\caption{ \label{FigureNc_vs_tc} The critical particle numbers $\tilde{N}_c$ for different values of $t_1 =q/n_1|\cbar{1}|$ calculated from the numerical solution of GP equation (orange) vs. from variational calculation with Gaussian ansatz (blue). The density at the center of droplet decreases with $q$; the kinetic energy and LHY pressure eventually overcomes MF attraction. 
}
\end{figure}

We numerically solve the modified GP equation \eqref{modifiedGPSpinor} to obtain the ground state wavefunctions for various values of the quadratic Zeeman coupling $q$ and the total number of particles $\tilde{N}$ using imaginary time propagation. 
Our results are shown in Fig.(\ref{FigureDropletWavefunctions}). 
The left panel in Fig.(\ref{FigureDropletWavefunctions}) shows the wavefunctions for various values of $\tilde{N}$ for fixed $q=0$. 
The polar spinor gas forms a self-bound quantum droplet for particle numbers above the critical value near $\tilde{N}_c \approx 19$ as in the Bose-Bose mixtures \cite{2015_Petrov_PRL}. 
Below this critical level $\tilde{N}_c$, the repulsive pressure due to the kinetic energy dominates the MF attraction and causes the droplet to expand to infinity. 
The right panel in  Fig.(\ref{FigureDropletWavefunctions}) shows the ground state wavefunctions for various values of $q$,  or $t$, for fixed $\tilde{N}= 500$. 
One can see that increasing $q$ strengthens the LHY energy of the spin fluctuations, which in turn decreases the maximum density of the droplet.
Above a critical value of the Zeeman energy, $q_c\approx 1.5$,
repulsion from the LHY energy combined with the kinetic energy quantum pressure overwhelms the MF attraction and the droplet again expands to infinity. Then, the quadratic Zeeman coupling can be used to tune the particle density at the center of the droplet.
However, a finite Zeeman energy, $q\neq 0$, also increases the critical value of $N_c$. Since the droplet wavefunction is Gaussian to a good accuracy around the critical region, we study the $q_c$ vs. $N_c$ using a Gaussian ansatz and compare with the numerical solution for different values of $t_1=q/n_1|\cbar{1}|$. 
As shown in Fig.(\ref{FigureNc_vs_tc}), $N_c$ increases with $t_1$ monotonically.

Polar droplet discussion has so far considered the parameter region in which $c_1>0$ and $q>0$. However,  the ground state also has polar order with ferromagnetic coupling $c_1<0$  and strong Zeeman shift $q>-2nc_1$ (blue region within the second quadrant in Fig. \ref{fig:phase_diagram}). In this regime, the LHY energy is obtained by replacing $I(t)$ with $I(t-2)$ in all of our previous calculations. We find that droplet formation is possible for all the parameter regimes with a polar ground state. 

\section{\label{sec:AF Spin-1 Droplet} Anti-Ferromagnetic Spin-1 Droplet}

Antiferromagnetic phase realized for $c_1>0$ and $q<0$ with an order parameter $\tau^{AF} = (1 \ 0 \ 1)/\sqrt{2}$ has the following total GS energy including the LHY correction: 
\begin{eqnarray}
\frac{E_0^{AF}}{V} &&= qn + \frac{n^2c_0}{2} \left[ 1 + 
\alpha\sqrt{nc_0^3}\right] \nonumber \\ 
&&+ 
\frac{\alpha n^2|c_1|}{2}
\sqrt{n|c_1|^3} \left[1+ I(t)\right],
\label{AFGroundState}
\end{eqnarray}
where $t = |q|/nc_1\equiv|q|/n|c_1|$ since $c_1>0$ here. 
Following the same steps given in Sec.~\ref{sec: Polar Spin-1 Droplet}, we consider the collapse induced in the MF level with $c_0<0$ and the stabilization with the fluctuations in the spin channel, and ignore the soft mode fluctuations in the density channel.
The pressure is calculated similarly as
\begin{eqnarray}
P &&= \frac{n^2c_0}{2} + 
\frac{\alpha n^2|c_1|}{4} 
\sqrt{n|c_1|^3}\  g(t)
\end{eqnarray}
where $g(t) \equiv 3 + 3I(t) -2tI'(t)$. 
The equilibrium density for the antiferromagnetic droplet is obtained from vanishing pressure as 
\begin{eqnarray}
n_{0} &&= \frac{4|\cbar{0}|^2}{\alpha^2 \cbar{1}^5 \ g^2(t_0) }
\end{eqnarray}
where $t_0 = |q|/n_0\cbar{1}$. The equilibrium density is positive and finite, since $g(t)$ is a monotonically increasing function of $t$, for $t\ge 0$.

The total energy functional in the antiferromagnetic phase with wavefunction $\Psi(\mathbf{r})=\psi(\mathbf{r})\tau^{AF}$
can be written as
\begin{eqnarray}
\mathcal{E}[\psi^*,\psi] &=& \frac{\hbar^2}{2M}|\nabla\psi|^2 + q|\psi|^2
+\frac{c_0}{2}|\psi|^4 
\nonumber\\
&+&\frac{\alpha}{2} |c_1|^{5/2}|\psi|^5
\left(1+
I\left[\frac{q}{|c_1||\psi|^2}\right]
\right)
\end{eqnarray}
We express the droplet wavefunction as 
$\psi(\mathbf{r}) = \sqrt{n_1} \phi(\mathbf{r})$ where $n_1 
= |\cbar{0}|^2/9\alpha^2 |\cbar{1}|^5$
is obtained from $n_0$ by taking the limit $q\rightarrow 0$. Defining $t_1=|q|/n_1c_1$ similarly, the variational minimization of energy in the grand canonical ensemble gives
the following modified GP equation 
\begin{eqnarray}
\tilde{\mu}\phi &&=
-\frac{1}{2}\tilde{\nabla}^2\phi 
+3t_1\frac{|c_1|}{|c_0|}\phi
-3|\phi|^2\phi \\
&&+\left\{ \frac{5}{4} \bigg[ 1 + I\left( \frac{t_1}{|\phi|^2} \right) \bigg]  |\phi|^3 -\frac{1}{2} I'  \left( \frac{t_1}{|\phi|^2} \right) t_1 |\phi| \right\} \phi.\nonumber 
\label{modifiedGPSpinorAF}
\end{eqnarray}
This equation of motion also reduces to the same form with Eq.~\eqref{modifiedGPSpinor}, when $q\rightarrow 0$, i.e. $t_1 \rightarrow 0$. Even though the AF phase yields the same ground-state wavefunctions with the polar case when $q=0$, the effect of non-zero $|q|$ values are different for AF and polar cases. The AF phase stabilizes the droplet with contributions from both density and spin fluctuations, while the spin fluctuation is the only stabilizing mechanism in the polar case. 

\begin{figure}[h]
\includegraphics[width=0.4\textwidth]{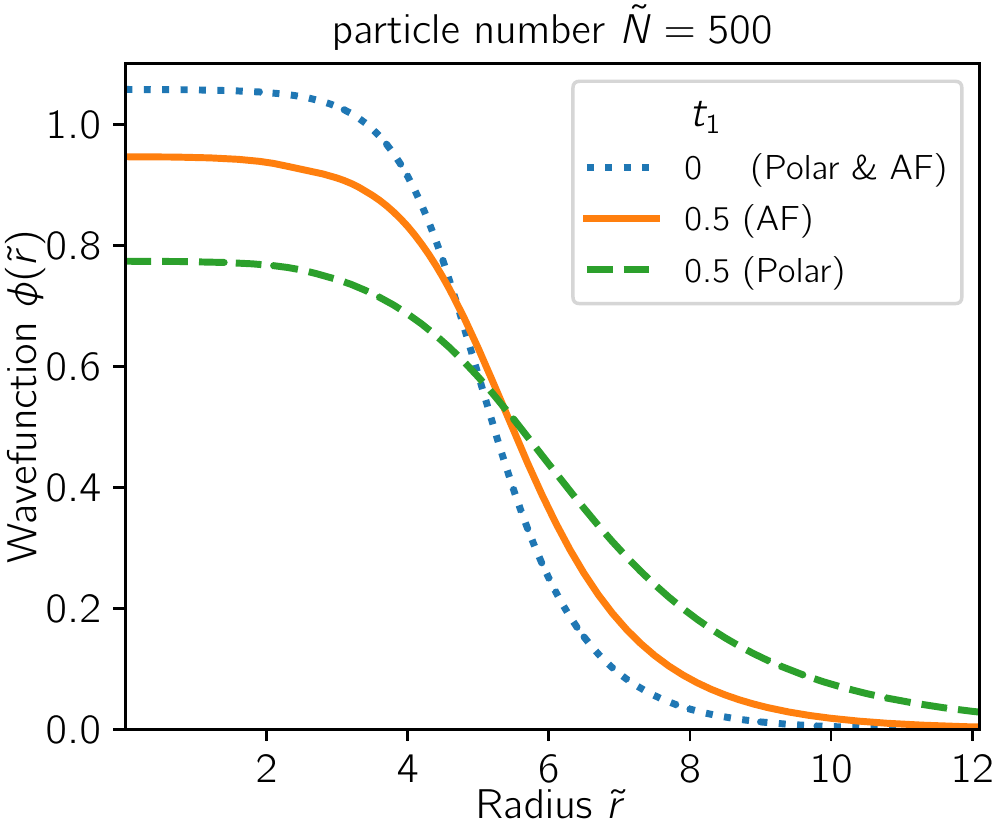}
\caption{ \label{DropletWFs_AF} Comparison of the ground state droplet wavefunctions for $t_1=-0.5,0,0.5$ with the particle number $\tilde{N} = 500$. The LHY energy dependence on the quadratic Zeeman shift depends on the order parameter. }
\end{figure}

To investigate the stability of a finite size AF spinor droplet, we solve the modified GP equation numerically and calculate the ground state wavefunctions in Fig.(\ref{DropletWFs_AF}). The stronger quadratic Zeeman shift again implies less density for the droplet within the central region. A further increase in $|q|$ causes the expansion of the droplet similar to the polar phase. Hence, the quadratic Zeeman effect can again be used to control the density of the droplet.    

\section{\label{sec:Experimental Discussion and Conclusion} Discussion of experimental realization and Conclusion}

The phase diagram and the stability of spin-1 Bose is controlled by three parameters: density-density interaction  $\cbar{0}$, spin-spin interaction $\cbar{1}$, and quadratic Zeeman energy $q$. A self-trapped droplet forms when the MF instability towards mechanical collapse induced with $\cbar{0} < 0 $ is balanced with the LHY quantum fluctuations. The derivation of the LHY energy is based on a perturbative expansion within the Bogoliubov theory and requires diluteness, which is satisfied if $|\cbar{0}|/|\cbar{1}|$ is sufficiently small. In the case of a finite droplet, additional quantum pressure coming from kinetic energy, the total particle number constraint, and the quadratic Zeeman shift give rise to a rich but complex interplay to determine the full stability. 

Experimentally, the spinor BECs obtained so far are not favorable for droplet formation since they are all mechanically stable with $\cbar{0}>0$ \cite{1998_Nature_Ketterle_Spinor,2004_PRL_Chapman_spinor,2004_PRA_Hirano_MagneticControl_q,2006_PRA_Petit_Spinor,2011_PRA_Kurn_Spinor,2013_RMP_Spinor_Kurn_Ueda}. For the most commonly used alkali atoms Na and Rb, $\cbar{0}$ is an order of magnitude larger than $|\cbar{1}|$ \cite{2013_RMP_Spinor_Kurn_Ueda},  while for Lithium $|\cbar{0}|/|\cbar{1}| \sim 0.46$ \cite{2020_PRR_Choi}.  To our knowledge, there is no fundamental reason for $\cbar{0}$ to be positive or much larger than $\cbar{1}$ in a complicated atom-atom scattering process, and favorable parameters may emerge in a novel hyperfine manifold as more atomic gases are trapped and cooled with new species and mixtures. Furthermore, the spinor gases lack the standard magnetic Feshbach resonance for tuning the interaction strength but there are theory proposals for scattering length tuning with optical Feshbach resonances which may soon be experimentally realized \cite{2015_PRA_Optical_Feschbach_Resonance,2018_Nature_Ott_Optical_Feshbach_Resonance}. We want to stress that the required change for the gas to go into the self-trapped droplet regime is not extraordinarily large. Hypothetically, consider an atom with 0-channel and 2-channel scattering lengths $a_0 = -100a_B$ and $a_2 = +45a_B$, which yields $\cbar{0} = \frac{4\pi \hbar^2}{M}(-\frac{10}{3} a_B)<0$ and  $\cbar{1} = \frac{4\pi \hbar^2}{M}(\frac{145}{3} a_B)>0$ with the ratio $\frac{\cbar{0}}{\cbar{1}} \simeq -0.07 $. For atom with 23 amu mass, the expected density at the center of a saturated droplet is $n_1 \approx 13\times 10^{14} cm^{-3}$ when $q=0$ and  a droplet with $N=25,000$ particles would be self trapped. The correlation length becomes $\xi \simeq 1 \mu m$. For total number of particles  $130,000 \ (\tilde{N} = 100)$, the droplet will be stable up to  $t_1 \simeq 1 $, thus the quadratic Zeeman shift can be varied between 0 to 20 $kHz$.   The droplet size is in the  $3-6 \mu m$ interval for $N \approx 25,000-150,000$ particles.

In conclusion, similar to the droplet formations in the dipolar and binary mixture gases, we predict a self-trapped droplet for the spin-1 gas in the polar and anti-ferromagnetic phases. The mechanism behind this stability is the competition between the MF attraction and LHY repulsion induced by spin fluctuations. The quadratic Zeeman effect can be used to control the stability of the droplet and its density. We hope that parameters favorable to droplet formation can be experimentally realized in new cold atom species or by adjusting the scattering lengths. Our results can be extended to non-zero magnetization, spin-2 gases, and spinor mixtures. Furthermore, other manifestations of the beyond MF interactions within the spinor BECs may also provide an exciting research direction.


\appendix*

\section{Analytical approximation for $I(t)$}

We use a change of variable variable $y \equiv x^2 + t + 1$ in
the integral 
\eqref{polarGroundState} 
and expand $\sqrt{1-1/y^2}$ in Taylor series up to the second order 
in the domain $x\ge 0$ and $t \ge 0$ and obtain
\begin{eqnarray}
I(t)  &&=  -\frac{15}{16\sqrt 2} \int_0^{\infty} dx  \left(\frac{-(t+1)}{x^2 + t +1} + \frac{x^2}{4(x^2 + t +1)^3}\right). \nonumber 
\label{I1_numerical_Appendix3}
\end{eqnarray}
Each term above can be calculated to give:
\begin{eqnarray}
I(t)  \approx \frac{15\pi}{32\sqrt{2}} \left[ \sqrt{t+1} - \frac{1}{32} \frac{1}{(t+1)^{3/2}} \right].
\label{I1_analytical_Appendix} 
\end{eqnarray}
Higher order terms in the expansion of $\sqrt{1-1/y^2}$ can improve the accuracy but we numerically checked that the second order expansion is sufficient up to less than one percent error for all $t$ values.
\acknowledgments

This work is supported by TUBITAK 2236 Co-funded
Brain Circulation Scheme 2 (CoCirculation2) Project No.
120C066 (A.K.). 

%

\end{document}